# Learning to see through multimode fibers

Navid Borhani,[1#] Eirini Kakkava,[1#] Christophe Moser,[2] Demetri Psaltis[1*]

*1 Optics Laboratory, School of Engineering, Ecole Polytechnique Fédérale De Lausanne, Lausanne, Switzerland*

*2 Laboratory of Applied Photonic Devices, School of Engineering, Ecole Polytechnique Fédérale De Lausanne, Lausanne, Switzerland*

*\*Corresponding author: demetri.psaltis@epfl.ch*

*#These authors contributed equally to this work*

## Abstract

We use Deep Neural Networks (DNNs) to classify and reconstruct a large database of handwritten digits from the intensity of the speckle patterns that result after the images propagated through multimode fibers (MMF). Images transmitted through fibers with up to 1km length were recovered. The ability of the network to recognize the input degraded with fiber length but the performance could be enhanced if the neural networks were trained to first reconstruct the image and then classify it rather than classify it directly from the speckle intensity.

## 1. INTRODUCTION

Optical fibers are prominently used in telecommunications [1] and endoscopy for medical diagnosis [2]. The need to increase the information throughput has encouraged the consideration of transmitting information through parallel channels in multimode fibers (MMFs). In telecommunications, this translates to launching multiple channels with different spatial modes. For endoscopy applications, the spatial modes of the MMF are used to carry the information of the different pixels in the image. However, when a pattern is projected on the proximal side of a MMF, the image we get on the distal side is a speckle pattern since the input couples into multiple fiber modes, which travel with different propagation constants along the fiber length. Additionally, local defects along the fiber length induce mode coupling, which further randomizes the propagation of the input field. Therefore, the phase between local image features decorrelates fast after a few millimeters of a MMF, resulting in the formation of a speckle pattern.

The transmission of images through a MMF was first presented in 1967 by Spitz and Wertz [3] who demonstrated experimentally that the distortion introduced by modal dispersion can be undone by phase conjugation. In the intervening years numerous publications have described methods for transmitting image information through MMFs or scattering media [4–15]. These methods rely on coherent (holographic) recording of the speckle pattern detected at the distal end of the fiber and they use phase conjugation or the transmission matrix method to compensate for the effects of modal dispersion and either focus the light at the distal end or project focused images. Iterative optimization algorithms can also optimize the phase of the input field in the MMFs in order to obtain the desired output [16]. There were two reports on artificial neural networks (ANNs) implemented for recovering the images transmitted through MMFs [17,18]. In these early demonstrations, 2-layer networks were trained and were able to recognize a few (~10) images.

Learning techniques have been employed in a variety of ways in optical systems. Examples have been used to train a reconfigurable optical systems [19–22] or train a digital computer to interpret or control the operation of a fixed optical system [23–25]. The work presented in this paper belongs to the second category. We collect a large number of intensity speckle patterns produced by launching images through a MMF and use these examples to train a DNN to interpret the input to the fiber. Specifically, we demonstrate the use of modern deep neural networks architectures [24,25] with up to 14 hidden layers which were trained on a database of 20,000 handwritten digits. Recognition or reproduction of an image launched at the proximal end of the fiber was achieved by detecting only the light intensity at the distal end facet. Then, we show that the classification performance by the DNN is greatly enhanced if we first use the network to reconstruct the input to the fiber followed by a separate DNN that is trained to recognize the reconstructed input images.

In the following section, we describe the experimental apparatus used to collect the database with which we trained the DNN to recognize the digits presented at the input. The same dataset was used to train a different DNN to reconstruct the input digits given the speckle pattern measured at the distal end. The performance was evaluated for different fiber lengths up to a maximum of 1km. We conclude with a discussion about the relative merits of using a combination of intensity detection with a DNN to interpret the measured data versus coherent (holographic) recording and linear inverse scattering methods to retrieve the input to the fiber.

## 2. MATERIALS AND METHODS

### A. Experimental setup

The optical system used to collect the data is shown in Fig. 1. The laser beam of a 560nm wavelength diode laser is used to illuminate a graded-index (GRIN) multimode fiber with 62.5μm core diameter and numerical aperture (NA) of 0.275 (GIF625, Thorlabs). The fiber supports approximately 4500 spatial modes at the specific wavelength. The input patterns are displayed on a spatial light modulator (SLM, 1920x1080 pixels, Pluto-Vis, Holoeye) and the SLM plane is imaged onto the proximal facet of the MMF by means of a 4f imaging system. Another 4f system is placed at the distal end of the fiber to image the speckle pattern emerging from the distal facet on a CCD camera (Chameleon 3, 1024x1280 pixels, Mono, Point Grey). An additional camera is used on the proximal side to monitor the images reflected by the SLM. A halfwave plate and a linear polarizer are placed before and after the SLM (see Fig.1) respectively in order to test both phase and amplitude patterns as inputs to the GRIN fiber.

In our experiments, the patterns generated by the SLM were handwritten digits from the MNIST database. Before processed by the DNN, each image recorded by CCD1 or CCD2 is cropped to a 1024x1024 pixels window centered on the digit and the speckle respectively. The cropped images were then downsampled to 32x32 pixels and used as input for the DNNs. An example of the projected digits at the proximal fiber facet is shown in Fig. 2, where the digits zero and four are shown for both amplitude (Fig. 2c-d) and phase modulation (Fig. 2e-f) along with the corresponding speckle patterns captured at the distal fiber end for a GRIN fiber that is 2cm long. The speckle patterns (Fig. 2g-h) look similar to one another because their appearance is dominated by the DC component of the light from the SLM. However, when we subtract the intensity patterns (Figs. 2d and 2h) corresponding to the two digits, we reveal that there is a significant difference (Fig. 2i) which can be picked up by the DNN to distinguish the two inputs. The results presented in the remainder of the paper are obtained by adjusting the SLM so that the patterns entering the fiber are phase only or amplitude modulated images of the digits.

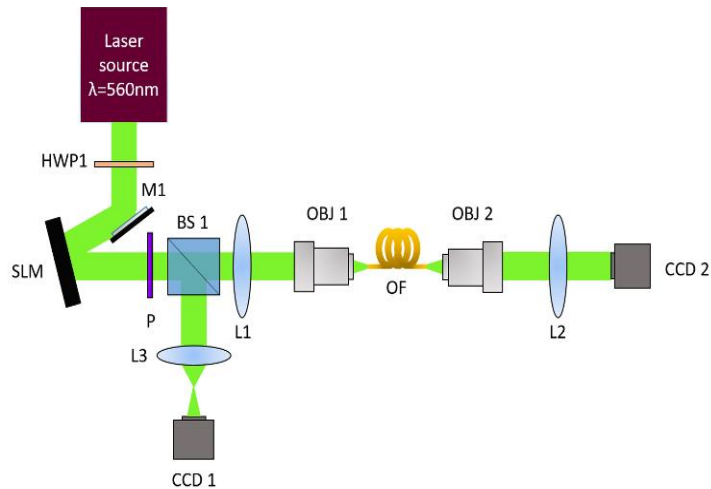

**Figure 1.** The experimental setup for pattern transmission through the MMF. The laser beam is expanded and collimated and is directed onto the SLM. The light is modulated by the SLM and its plane is imaged by means of a 4f system onto the proximal facet of a GRIN fiber. The distal facet is imaged by a second 4f system on a CCD camera (CCD2). The modulated light after the SLM is also captured by an imaging system in 2f configuration on the CCD1. HWP: halfwave plate, M:mirror, SLM: spatial light modulator, P: linear polarizer, L: lens, BS: beam splitter, OBJ: microscope objective lens, OF: optical fiber, CCD: camera.

## B. Data processing

A 'VGG' type convolutional neural network (CNN), as developed by Simonyan & Zisserman [26] was used to classify the distal speckle images and reconstructed SLM input images (Fig. 3a). These networks consist of a convolutional front end with downsampling for encoding, and a fully connected back end for classification; see Fig. 3a for details. The use of such deep CNN with very small filter kernels has been shown to provide high image classification accuracies.

A 'U-net' type CNN with 14 hidden layers, as developed by Ronneberger et al. [27], was used to reconstruct the SLM input image from the recorded distal speckle intensity pattern (Fig. 3b). This nearly symmetric network architecture comprises a convolutional encoding frontend with downsampling to capture context, and a deconvolutional decoding backend with upsampling for localization; see Fig.3b for details. Skip connections copy feature layers produced in the contracting path with features layers in the expanding path of the same size, thus improving localization.

For training both networks, the obtained 20k distal speckle pattern images were randomly split into 16k training, 2k validation, and 2k testing sets. The training sets were processed in 50 and 500 image batches for the reconstruction and classification networks, respectively, with batch shuffling to minimize over fitting. An Adam optimizer with a learning rate of $1\times10^{-4}$ was used to minimize a mean square error cost function. The networks were trained for a maximum of 50 epochs. For each case, training was carried out 10 times to provide statistics for the training accuracies.

The DNNs were implemented using the TensorFlow 1.5 Python library on a single NVIDIA GeForce GTX 1080Ti graphics processing unit.

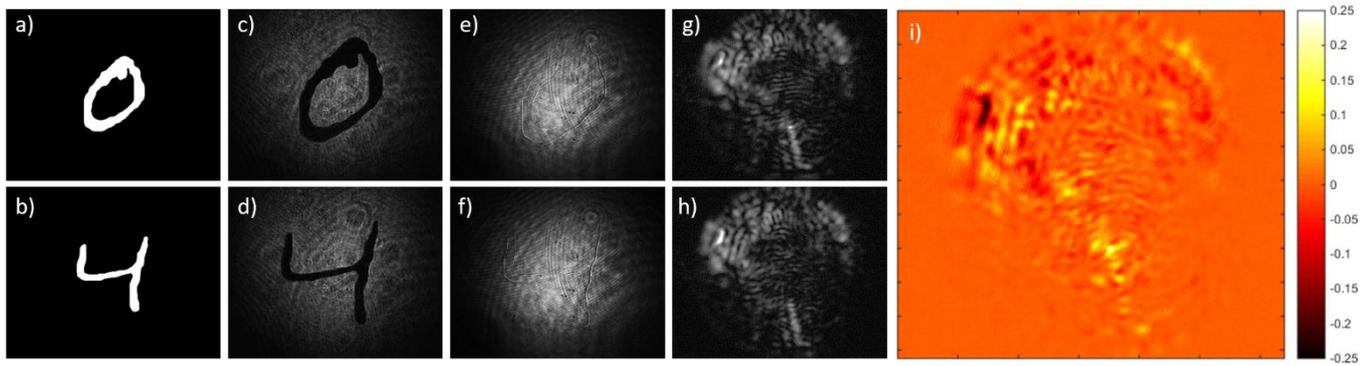

**Figure 2.** Images of the digits 0 and 4: a-b) input pattern on the SLM, c-d) amplitude modulated output from the SLM, e-f) phase modulated output from the SLM, g-h) speckle patterns of each digit respectively for the amplitude inputs and i) the difference between speckles g and h.

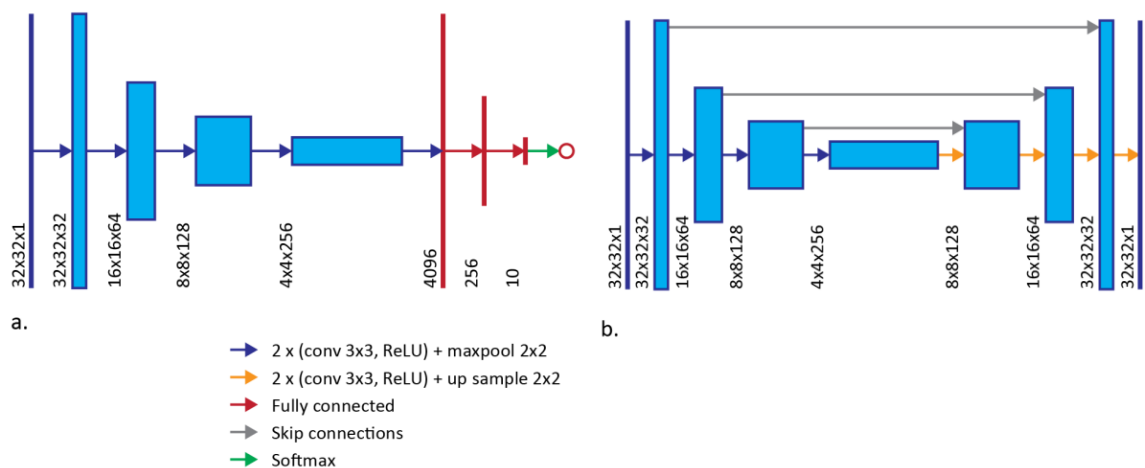

**Figure 3.** Details of the implemented (a) VGG type image classifier, and (b) U-net type image reconstruction convolutional neural networks.

# 3. RESULTS

## A. Input reconstruction

In a first step, the ability of our DNN to reconstruct the input digits from the distal speckle intensity patterns was tested. In Fig. 4, we present the results of the reconstruction for the 0.1m, 10m, and 1km fiber lengths with amplitude modulated inputs into the proximal facet of the GRIN fiber.

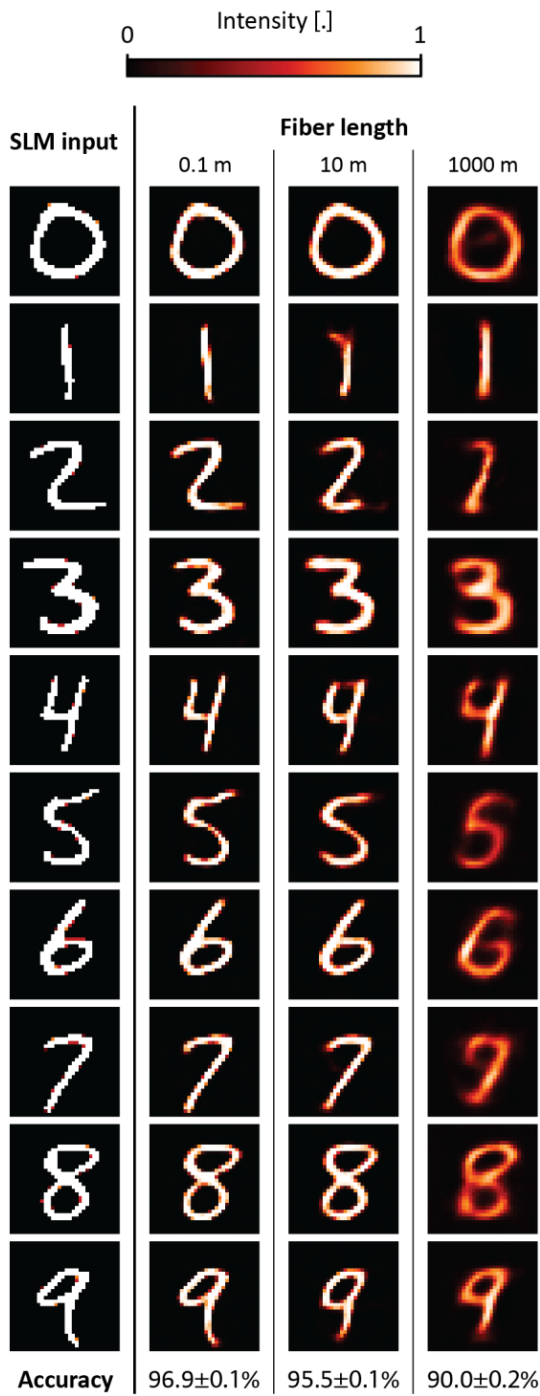

**Figure 4.** Examples and accuracies of the reconstructed SLM input images from the recorded distal speckle intensity patterns for amplitude modulated proximal inputs

Although appearing random, the speckle patterns contain information about the propagation of the input field through the fiber. In fact, the results confirm the above statement, showing that the recovery of the input is possible with an intensity-only image of the distal speckle pattern using the U-net CNN. Based on the reconstructed images obtained from our experiments (Fig.4), the fidelity of the reconstruction decreases from 97.6% for a 0.1m fiber to 90.0% for an 1km fiber.

In the case of the 1km long GRIN fiber, the speckle pattern at the distal end was unstable. Local temperature nonuniformities in the fiber induce changes in the optical path, due to both thermal expansion of the material and change of its refractive index. Thermal convection around the fiber can lead to drifting of the distal speckle pattern in time creating an extra "noise" on the acquired speckle patterns acquired (Supplementary, see Visualization 1). Therefore, further care could be taken to thermally isolate the fiber and to maintain an isothermal environment, which might give an increase of performance. The high fidelity of the reconstructed SLM input images also show that this technique effectively denoises the system by removing artefacts associated with the optical setup. Fidelity was measured as the percent mean square error of the reconstruction compared to the input. For example, the network recovers the SLM input image shown in Fig. 2a from the distal speckle intensity pattern shown in Fig. 2g, while eliminating artifacts projected onto the proximal facet of the fiber, as shown in Figs. 2c & 2e, as well as subsequent artifacts due to flaws and dirt or even misalignments on the proximal facet of the fiber.

In the movie, included in the online version of the (Visualisation 1), one can observe changes in the distal speckle pattern over a 5s period. This movie was recorded at a frame rate of 83fps for the 1km long GRIN fiber with a constant blank image as the SLM input. As shown, although the proximal input does not change, the speckle intensity at the distal end of the fiber changes rapidly with time. This can be attributed to fluctuations of the ambient temperature or airflow over the optical setup that induce slight perturbations on the GRIN fiber that become significant over its 1km length. Therefore, changes on the distal output caused by the projection of different digits while the training dataset is acquired can be buried in the "noise" caused by the drifting of the speckle pattern. In order to test the effect of the drifting distal speckle patterns on the accuracy of the classifications, the VGG network was trained on the first 10,000 samples of images of an acquired dataset and tested on images from the second half (recorded several hours later); and vice versa. The results showed no significant change in the classification accuracy. This suggests that the fluctuations seen in the video are not entirely random and the neural network has learned them.

**B. Input classification**

Results for the classification of the distal speckle intensity patterns are presented in Table 1 and Fig.5. These show that the classification accuracy, defined as the percentage of correctly recognized digits, decreases with increasing fiber length for both amplitude and phase modulated proximal facet input modes. Generally, the accuracy decreases from 90% for a 2cm fiber to 30% for a 1km fiber. This compares with a classification accuracy of 98.4% for the original SLM input digit images. This decrease can be attributed to increased scattering losses, mode coupling, and drifting of the distal speckle pattern with increasing fiber lengths. The results also show that phase modulated input provides slightly better classification accuracies probably due to the more uniform distribution of the injected light across the fiber modes.

**Table 1. Classification accuracy for the four different fiber lengths using amplitude or phase input patterns. Classification was carried out on either the intensity image of the distal speckle patterns or on the reconstructed SLM inputs.**

| Fiber length [m] | Proximal input | Classification accuracy [%] | |
|---|---|---|---|
| | | From distal speckle intensity | From reconstructed input |
| 0.02 | Amplitude | 92.7 ± 0.5 | 98.1 ± 0.4 |
| | Phase | 95.1 ± 0.6 | 98.1 ± 0.3 |
| 0.1 | Amplitude | 90.7 ± 0.8 | 97.5 ± 0.5 |
| | Phase | 92.2 ± 0.7 | 97.5 ± 0.3 |
| 10 | Amplitude | 81.9 ± 1.6 | 96.5 ± 0.4 |
| | Phase | 87.2 ± 0.9 | 96.8 ± 0.5 |
| 1000 | Amplitude | 29.3 ± 5.5 | 69.9 ± 0.9 |
| | Phase | 22.4 ± 2.2 | 57.0 ± 1.0 |

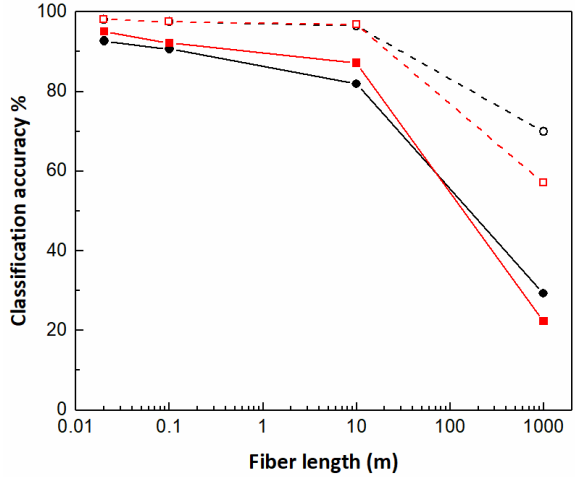

**Figure 5.** Classification accuracy as a function of fiber length. (Solid line: The inputs to the VGG-CNN are the recorded speckles at the distal fiber end, Dotted line: The inputs to the VGG-CNN are the reconstructed images obtained by the U-net-CNN, Circles: Amplitude images projected on the proximal fiber facet, Squares: Phase images projected on the proximal fiber facet)

In order to improve the classification accuracies, the neural network was also trained with the reconstructed SLM input images. As shown in Table 1 and by the normalized confusion matrices in Fig. 6, this provided a significant increase in classification accuracy. For the 1km case, there is a general confusion between the 4 and 9s digits, and between the 3, 5, 6, and 8's. The similarities between these classes are also evident in the reconstructed SLM input images for the 1km fiber shown in Fig. 4.

## 4. DISCUSSION

The main result of this paper is that DNN's can efficiently reconstruct and recognize the inputs to a MMF from intensity only measurements of its corresponding output. The measured classification accuracy was excellent for the 10m fiber (96.8%) and reduced to 69.9% for the 1km fiber. In all cases, the classification performance improved when we first used the U-net DNN to reconstruct the input image followed by a second DNN (VGG) that was trained to classify the reconstructed images. We attribute this to the fact that for the longer fiber the mapping from input to output becomes a random mapping [28] and objects that are similar to one-another at the input are dispersed in the intensity measurement at the distal end. Therefore, the ability of the classifier-DNN to generalize (recognize objects it has not seen before) diminishes for longer fibers. When we first recover the input images with a U-net DNN, the random mapping is partially inverted and the classification network can recognize objects of the same class it has not seen before. This behavior is evidenced in Figure 7 where the classification performance of the VGG network when trained with the intensity of the raw speckle patterns is plotted as a function of iteration number during the learning process for the 10m (Figure 7a) and the 1km (Fig 7c) fibers. For the 10m fiber the classification accuracy is the same for the training and validation sets. On the contrary for the 1km fiber, in steady state, the training set is memorized well but the validation set is classified accurately only 29.3 % of the time. The discrepancy in recognition rate between the training set and validation sets is an indication that the network is not able to generalize well. Also shown in Fig.7 are the learning curves when training the VGG network with the reconstructed images from the U-nets. In this case, the recognition rate is the same for the training and validation sets. In general, we can improve the recognition rate on the validation and test sets while decreasing the performance on the training set by reducing the number of weights in the network and/or increasing the size of the training set.

The recognition or reconstruction of the field at the proximal end of a MMF from complex field measurements at the distal end can be considered as an alternative to the DNN based inversion methods we describe in this paper. The simplicity of intensity only detection is a clear advantage in practice. At the same time, linear inversion methods (the transmission matrix) learn the fiber not the inputs. In other words, any input can be recognized or reproduced. DNNs on the other hand are trained on a class of objects and rely on statistical averaging within that class. In principle, the performance of the transmission matrix method should be independent of fiber length. However, as the fiber length increased additional background noise accumulates at the output because of scattering at the core for the fiber and the core-cladding interface. In addition, the temperature and mechanical instabilities that contaminate the measured data are to some extend learned by the DNN (Supplementary material, see

Visualization1) whereas they directly degrade the reconstructions of coherent methods. Finally, the neural network can be directly trained to reproduce or recognize the versions of the input images as they are stored in the computer. Any nonlinearities, aberrations, speckle, pixelation, phase wrapping, or other distortions that are introduced before the light enters the input facet of the MMF (i.e Figure 2a versus Figure 2b) are conveniently accounted for.

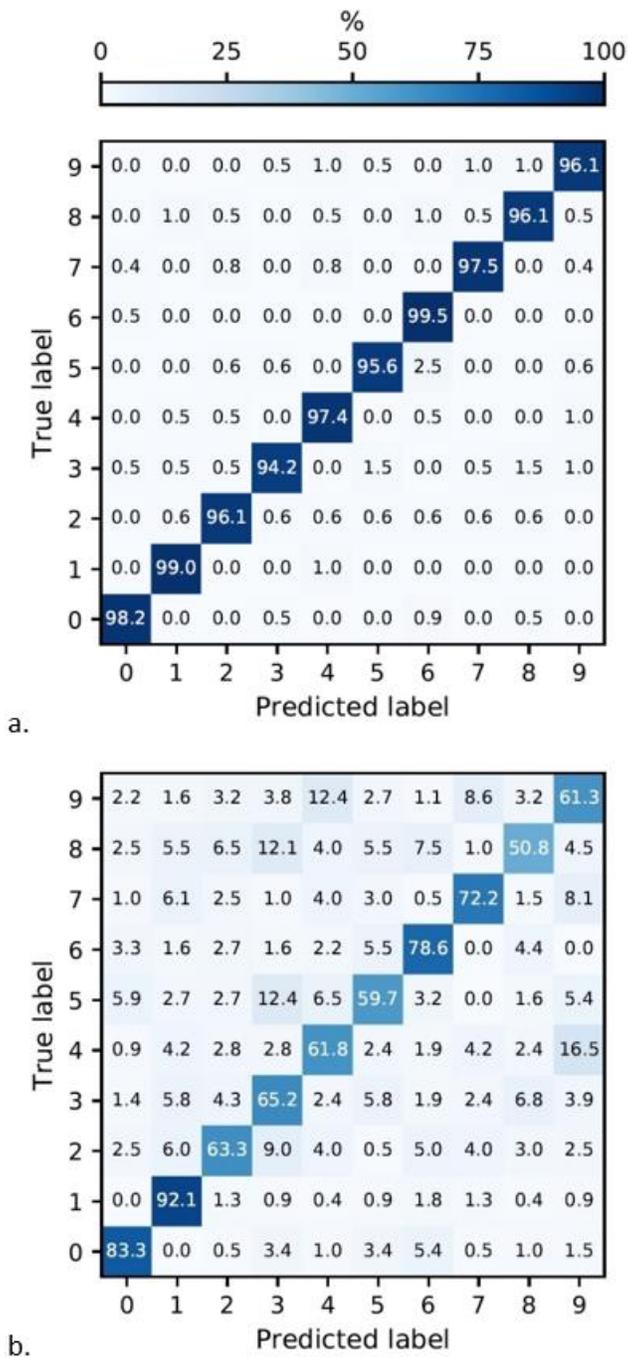

Figure 6. Normalized confusion matrices for the classification of the reconstructed SLM input images for the (a) 10m and (b) 1 km GRIN fiber with an amplitude modulated proximal input

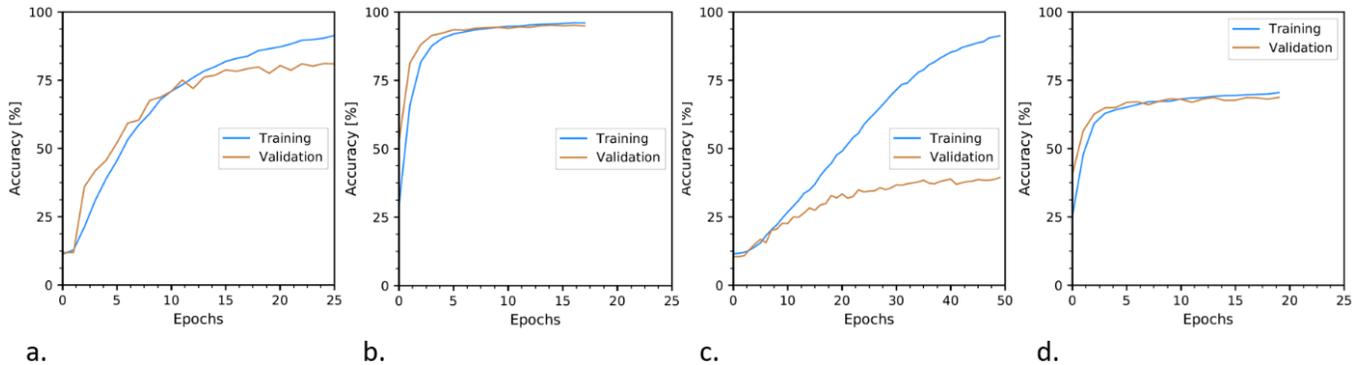

Figure 7. Training and validation classification accuracies as a function of epoch for the (a) 10m fiber distal speckle intensity pattern, (b) 10m fiber SLM reconstructed input, (c) 1km fiber distal speckle intensity pattern, and (d) 1km fiber SLM reconstructed input.

## 5. CONCLUSION

We have shown that DNNs can recognize and reconstruct distorted images at the output of a MMF. The DNNs can recover directly the phase of the input image. This is a doubly nonlinear mapping due to the square law at the output and the exponential dependence of the field in phase. An interesting subject to explore in the future is whether DNNs recognize inputs when the intensity of the light source is increased to the point where nonlinear effects significantly impact the output intensity patterns. A related question is whether neural network controllers [29] can be used to control nonlinear light propagation in MMF's [30,31].

**Funding.** This project was partially conducted with the support of the Swiss program: CEPF SFA, CERAMIC X.0: High-precision micro-manufacturing of ceramics and the Bertarelli Program in Translational Neuroscience and Neuroengineering (10271)

See also Supplementary material below for further information.

# Supplementary material: Learning to see through multimode fibers

Navid Borhani,[1#] Eirini Kakkava,[1#] Christophe Moser,[2] Demetri Psaltis[1*]

*1 Optics Laboratory, School of Engineering, Ecole Polytechnique Fédérale De Lausanne, Lausanne, Switzerland*

*2 Laboratory of Applied Photonic Devices, School of Engineering, Ecole Polytechnique Fédérale De Lausanne, Lausanne, Switzerland*

*Corresponding author: demetri.psaltis@epfl.ch

#These authors contributed equally to this work

## Abstract

In this document, we provide supplementary Figures of the measurements for each fiber length (2cm, 10cm, 10m and 1km) used in the work presented in the manuscript: "Learning to see through multimode fibers".

**1. SLM INPUT RECONSTRUCTION USING THE U-NET CNN**

**Figure S1.** Examples and accuracies of the reconstructed SLM input images from the recorded distal speckle intensity patterns.

Fig. S1 is the full version of Fig. 4 presented in the main manuscript. In this figure, we present examples of the reconstructed SLM input obtained by the U-net CNN trained on the recorded distal speckle intensities patterns. As well as results for the 2cm fiber, the case of phase modulated inputs are also shown. It is clear that the classification accuracy does not experience any deterioration for fibers up to 10m in length, thus proving that the use of DNNs can lead to new methods for image transmission through MMFs.

## 2. DATA CLASSIFICATION USING THE VGG CNN

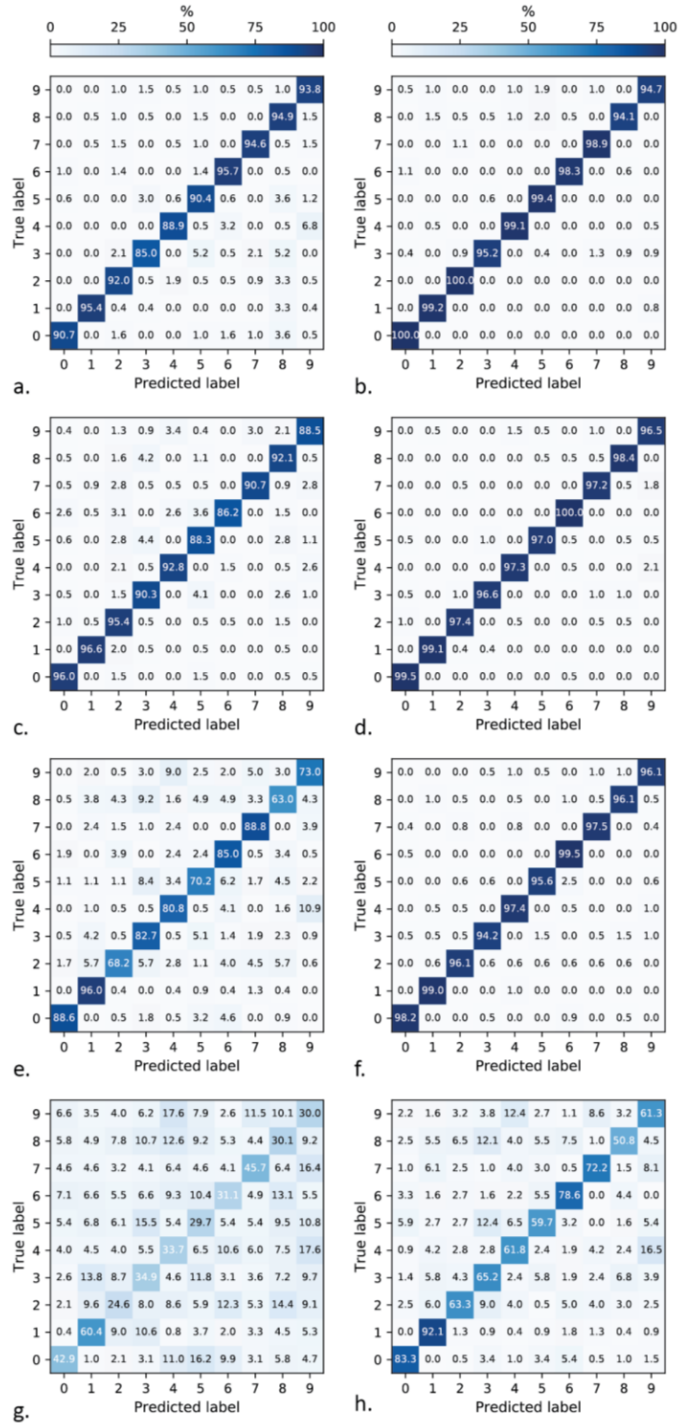

**Figure S2.** Normalized confusion matrices for classification with amplitude modulated proximal inputs for a) 2cm fiber speckle patterns, b) 2cm reconstructed SLM inputs, c) 10cm fiber speckle patterns, d) 10cm reconstructed SLM inputs, e) 10m fiber speckle patterns, f) 10m reconstructed SLM inputs, g) 1km fiber speckle patterns, and h) 1km reconstructed SLM inputs.

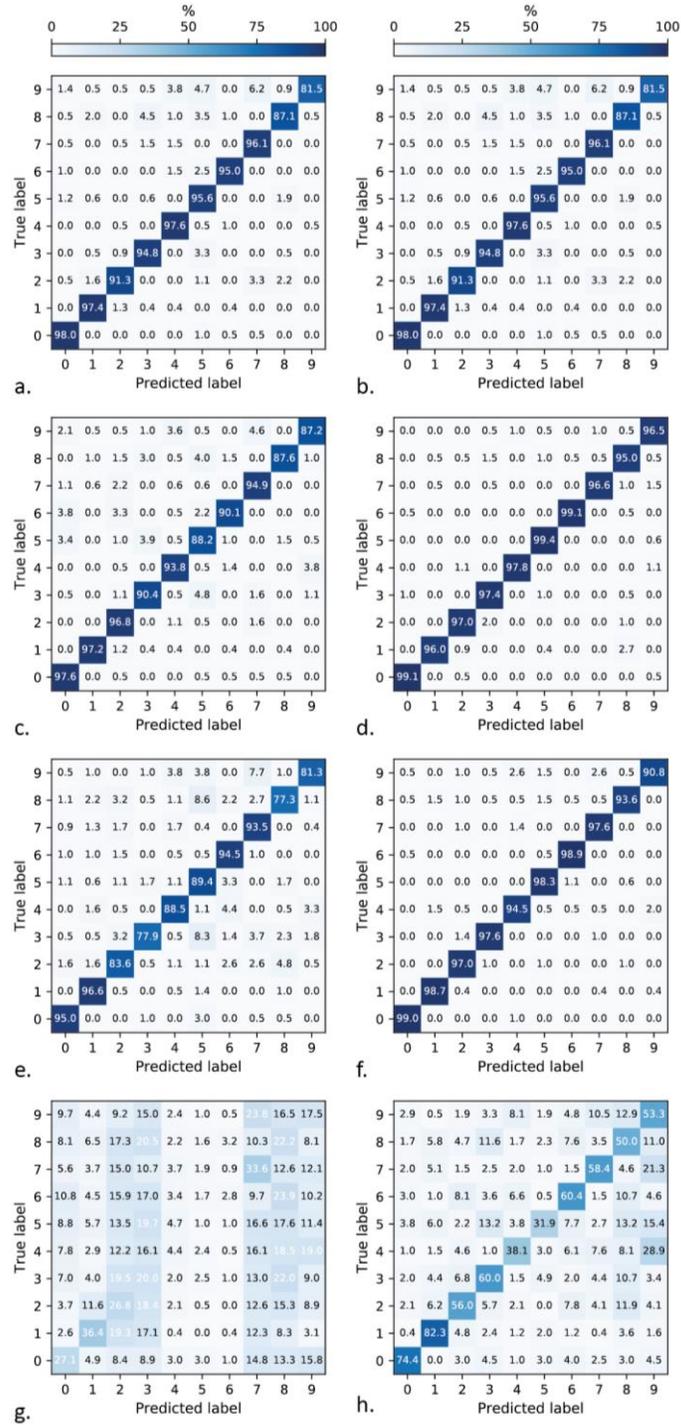

**Figure S3.** Normalized confusion matrices for classification with phase modulated proximal inputs for a) 2cm fiber speckle patterns, b) 2cm reconstructed SLM inputs, c) 10cm fiber speckle patterns, d) 10cm reconstructed SLM inputs, e) 10m fiber speckle patterns, f) 10m reconstructed SLM inputs, g) 1km fiber speckle patterns, and h) 1km reconstructed SLM inputs